\documentclass[aps,pra,floatfix,onecolumn,11pt,superscriptaddress,showkeys,a4paper]{revtex4}

\usepackage{amsmath}
\usepackage{amssymb}
\usepackage{latexsym}
\usepackage{amsfonts}

\usepackage{graphicx}
\usepackage{pdfpages}

\usepackage{mathptmx}

\setcitestyle{super}

\newcommand{\ket}[1]{\vert{#1}\rangle} 
\newcommand{\bra}[1]{\langle{#1}\vert} 
\newcommand{\bracket}[2]{\langle{#1}\vert{#2}\rangle} 
\newcommand{\proj}[1]{\ket{#1}\!\bra{#1}}

\newcommand{\one}{\openone}
\DeclareMathOperator{\Tr}{Tr}

\newcommand{\abs}[1]{\left|#1\right|} 

\newcommand{\beq}{\begin{equation}}
\newcommand{\eeq}{\end{equation}}
\newcommand{\up}{\uparrow}
\newcommand{\down}{\downarrow}

\begin{document}

\title{Open Quantum System Approach to the Modeling of Spin Recombination Reactions}

\author{M. Tiersch}
\email{markus.tiersch@oeaw.ac.at}
\affiliation{Institute for Quantum Optics and Quantum Information,
Austrian Academy of Sciences,
Technikerstr.~21A, A-6020 Innsbruck}
\affiliation{Institute for Theoretical Physics,
University of Innsbruck,
Technikerstr.~25, A-6020 Innsbruck, Austria}

\author{U. E. Steiner}
\affiliation{Fachbereich Chemie, Universit\"{a}t Konstanz, D-78457 Konstanz, Germany}

\author{S. Popescu}
\affiliation{H. H. Wills Physics Laboratory, University of Bristol, Tyndall Avenue, Bristol, BS8 1TL, United Kingdom}

\author{H. J. Briegel}
\affiliation{Institute for Quantum Optics and Quantum Information,
Austrian Academy of Sciences,
Technikerstr.~21A, A-6020 Innsbruck}
\affiliation{Institute for Theoretical Physics,
University of Innsbruck,
Technikerstr.~25, A-6020 Innsbruck, Austria}


\begin{abstract}
In theories of spin-dependent radical pair reactions, the time evolution of the radical pair, including the effect of the chemical kinetics, is described by a master equation in the Liouville formalism.
For the description of the chemical kinetics, a number of possible reaction operators have been formulated in the literature.
In this work, we present a framework that allows for a unified description of the various proposed mechanisms and the forms of reaction operators for the spin-selective recombination processes.
Based on the concept that master equations can be derived from a microscopic description of the spin system interacting with external degrees of freedom, it is possible to gain insight into the underlying microscopic processes and to develop a systematic approach towards determining the specific form of reaction operator in concrete scenarios.
\end{abstract}

\maketitle

\section{Introduction}

Spin chemistry is concerned with the interplay of chemical reaction kinetics and electron/nuclear spin dynamics.\cite{Salikhov,Steiner,Nagakura}
The central, though not exclusive, paradigm of spin chemistry is the radical pair mechanism.
It focuses on the chemical behavior of radical pairs, formed as reaction intermediates either as geminate radical pairs (RPs) in a correlated fashion from precursors of well-defined spin multiplicity, or in an uncorrelated fashion from the statistical encounter of free radicals (F-pairs).\cite{McLauchlan}
The key assumption in explaining spin chemical effects is the postulate of spin conservation during elementary chemical processes such as electron transfer or covalent bond cleavage and formation, an assumption that leads to a dependence of the reactive behavior of a RP on the different spin substates.
This notion implies that the relevant Hilbert space of the RP comprises a set of energetically close lying states differing essentially in electron/nuclear spin quantum numbers.
(It should be noted that in case of the electron spin it is usually an effective spin, comprising the different spin-orbit mixed character of the various substates of the RP considered.)
Insofar as the internal dynamics of this set of states can be described by an effective spin Hamiltonian, it is coherent.
Hence, in a kinetic theory, the RP has to be represented by a density matrix $\rho$ with the internal dynamics described by the commutator $[H,\rho]$.
However, regarding the correct description of the spin-selective chemical kinetics, the question arises how to account for it within the Liouville formalism.
Traditionally, in theoretical spin chemistry this has been done in an \emph{ad hoc} fashion, but taking into account the necessary conditions for conserving the properties of a density matrix (positive, Hermitian).
Very recently however, there has been a renewed debate about the correct way to treat the problem, with a number of different proposals~\cite{Kominis,Jones,Ilichov,Ivanov,Shushin,Jones2011,Kominis2011}, each starting with a different philosophy, using different formalisms and, more importantly, predicting different answers.
Which of them is correct and which are wrong? 
Here we show a framework that allows to treat all presented approaches on the same footing.
It demystifies the quantum measurement approach to spin chemistry, and provides a common basis, from which the various proposals can be obtained.
Thereby, we can identify valid approaches, and we provide a systematic way of constructing new ones.
The decision about which of the particular instances applies in a concrete scenario cannot be decided a-priori but depends on the precise physical situation.
In more detail, the situation is as follows.

For pedagogical reasons and to keep the analysis transparent, we will restrict ourselves to the case of a RP with exclusive singlet reactivity and, if concrete representations are given, we shall consider a simple two-level model ($T_0$, $S$) of the set of spin substates. 
The standard form of the reaction part of the stochastic Liouville equation (SLE) of the RP, as justified by the formal criterion that the density matrix must always remain positive definite, was given by Haberkorn~\cite{Haberkorn} as
\beq \label{eq:Haberkorn}
\left[\dot\rho\right]_\text{rxn} = -\frac{1}{2} k_S (Q_S \rho + \rho Q_S)
\eeq
where $Q_S$ is the projection operator on the singlet RP state:
\beq
Q_S=\proj{S}
\eeq
and $k_S$ is the specific rate constant of the singlet RP reaction.

Triggered by a paper of Kominis~\cite{Kominis} who first proposed to treat the spin selective reaction of RPs in the framework of quantum measurement theory, there has been a renewed debate on the physically correct form of the reaction operator.
Although the details of the formal derivation given by Kominis have been debated, the result
\beq \label{eq:Kominis}
\left[\dot\rho\right]_\text{rxn} = -\frac{1}{2} k_S (Q_S \rho + \rho Q_S) + k_S Q_S \rho Q_S
\eeq
did arouse some interest. The difference between \eqref{eq:Haberkorn} and \eqref{eq:Kominis} is that the latter conserves the trace of the density matrix.
It describes a mere singlet/triplet dephasing due to the coupling of the RP states to the reaction channel but does not account for the decay of the RP population, which Kominis took into account by a further \emph{ad hoc} assumption of a rate of $k_S \Tr\{Q_S \rho Q_S\}$ for the decay of RP population which, however, would not change the normalized density matrix of the RP.

Jones and Hore,~\cite{Jones} taking up the quantum measurement approach in a formally consistent way, arrived at the result:
\begin{align} \label{eq:JonesHore}
\left[\dot\rho\right]_\text{rxn}
&= -k_S \rho + k_S Q_T \rho Q_T \\
&= -\frac{1}{2} k_S (Q_S \rho + \rho Q_S) -\frac{1}{2} k_S (Q_S \rho + \rho Q_S) + k_S Q_S \rho Q_S. \nonumber
\end{align}
As indicated in the second line, using a resolution of the identity in the singlet/triplet subspace, $Q_S+Q_T=\one$ with $Q_T=\proj{T}$ being the projector on the triplet RP state, this expression comprises the Haberkorn type reaction operator (first term) plus, in the second and third term the dephasing operator as it occurred in~\eqref{eq:Kominis}, which is of Lindblad form.

\bigskip{}

\paragraph*{}
In follow-up work by Il'ichov and Anishchik,~\cite{Ilichov} Ivanov et al.,~\cite{Ivanov} and Shushin~\cite{Shushin}, it was attempted to derive appropriate forms of the reaction operator by starting out with explicit treatments of the interaction of the RP spin states with the environment, including the systems of intramolecular nuclear vibrations. All of these treatments essentially arrive at the Haberkorn reaction term and question the appropriateness of the quantum measurement-based view of the RP reaction.

In this work, we start from the same premise that a master equation description for the state evolution originates from microscopic dynamics of the considered quantum system interacting with additional, usually uncontrolled degrees of freedom named \emph{the environment}.
Although these degrees of freedom may allow for an approximate classical description, as a source of noise for example, a first-principle treatment should include them as quantum degrees of freedom.
Therefore, the evolution of the composite system, i.e.\ system \emph{and} environment, is coherent and described by a commutator that contains the system--environment interaction.
When restricting the formal description to the system only, i.e.\ treating the \emph{open} quantum system, this interaction results in an effective, in general non-unitary evolution that needs to be accommodated within the Liouville formalism, for example.
From this perspective it is possible to identify classes of system--environment interactions that lead to the specific proposed reaction operators, on one hand.
On the other hand, given that a master equation is an accurate description of the real reaction dynamics, this approach opens up new possibilities to identify the form of the reaction operator in concrete scenarios.

Within the chosen framework, i.e.\ including the environmental degrees of freedom into the description, it is furthermore possible to phrase the open system dynamics in terms of measurements.
Thereby we can embed the kinematics due to ``quantum measurements'' into a broader context, in which the quantum system of interest interacts with environmental degrees of freedom.
The effective dynamics of the system can always be interpreted as measurements by the environment.
Along the lines that ``any linear process affecting an open system A can be viewed as an unread generalized measurement, i.e.\ as resulting from an entanglement of A with an environment simulator B, followed by an unread projective measurement in B'' as has recently been stated by Haroche and Raimond~\cite{Haroche} and as it is known from measurement theory and open system dynamics\cite{vonNeumann,Kraus,Davies}, we will show in this work that a systematic view on the various approaches to the reaction term in spin chemistry can be obtained by taking into account how the environment is affected through the interaction with the RP.

We start out with defining dynamical maps that characterize in a very general way how the interaction may affect system and environment.
From these maps we derive pertinent Lindblad operators in the Hilbert space comprising the spin states of the RP and the reaction product.
Reducing these equations to the subspace of the RP leads to the various forms of master equations discussed in spin chemistry.
Finally, we indicate possible experimental avenues for the characterization of the reaction operator.

\section{Dynamical maps of system--environment interaction}

Before considering the dynamics, it is necessary to fix the concepts and to specify which degrees of freedom are considered.
The quantum system of interest are the spin degrees of freedom of the radical pair (RP).
The spins are usually the effective spins of two unpaired electrons, each being localized on its respective molecule of the RP.
In order to capture the full dynamics, it is necessary to take into account the different situations of whether or not the charges have recombined or a covalent bond has been formed.
We therefore describe the system Hilbert space by the spin state of the \emph{unrecombined} radical pair, which we limit for pedagogical reasons to the singlet state and the triplet state with zero projection of total spin, denoted by $\ket{S}$ and $\ket{T}$, respectively.
In addition, we include into the Hilbert space the \emph{recombined} state/reaction product state denoted by $\ket{P}$, which is also a singlet state but with a different electron configuration.
(Formally, we thus consider a system of two degrees of freedom: electron spin and electron configuration.
But in practice the reaction product will comprise an entire manifold of states orthogonal to the radical pair states $\ket{S}$ and $\ket{T}$. Since we do not intend to distinguish between different singlet reaction products, we can effectively lump together the entire reaction product manifold under the label $\ket{P}$.)

The environment degrees of freedom shall remain abstract for the moment, since it is not necessary to restrict the treatment to specific examples. For concreteness, the reader may, however, think of the state of the electromagnetic field, which may be in the vacuum state or with a photon emitted due to fluorescence in the course of recombination, or the relative position and orientation of the radical pair molecules carrying the unpaired electron spin, and the configuration of their solvent shells.

Let us first consider the interaction between the system and the environment for some time interval $\delta t$ under the interaction Hamiltonian $H_\text{int}$.
The quantum state evolution of system and environment can be captured completely by a map from all possible initial states to their respective final states. It is, however, sufficient to only consider all products of basis states of system and environment, thereby simply representing the matrix elements of the time-evolution operator $U(\delta t)=\exp(-iH_\text{int} \delta t)$.
The basis states of the two RP spins are the singlet state $\ket{S}_s$, and the triplet state with zero projection of spin, denoted by $\ket{T}_s$.
In addition to the basis states of the system, $\ket{S}_s$, $\ket{T}_s$, and $\ket{P}_s$, the environment, to which the system is coupled, is assumed to start from some initial state $\ket{0}_e$.
For clarity, system and environment states are labeled by subscript $s$ and $e$, respectively.

\subsection{Spontaneous decay}

The first example of a system--environment interaction is a straightforward realization of a recombination dynamics with a minimum number of terms, and we will recognize this to be a simple, spontaneous decay process.
It is described by the dynamical mapping:
\begin{align}
\ket{S}_s \ket{0}_e & \to a \ket{P}_s \ket{\sigma}_e + b \ket{S}_s \ket{0}_e, \label{eq:decayS} \\
\ket{T}_s \ket{0}_e & \to \phantom{a} \ket{T}_s \ket{0}_e. \label{eq:decayT}
\end{align}
The first part in \eqref{eq:decayS} captures the fact that with probability $\abs{a}^2$ an initial singlet spin state evolves into some reaction product state $\ket{P}_s$ with the environment evolving into some corresponding state $\ket{\sigma}_e$, e.g.\ with a photon emitted due to fluorescence.
Note, however, that the environment state $\ket{\sigma}_e$ might be some effective state of many degrees of freedom comprising the environment.
With probability $\abs{b}^2=1-\abs{a}^2$ both remain unaffected.
The triplet does not interact with the environment and hence both do not evolve under the interaction Hamiltonian.
The probability amplitudes $a$ and $b$ depend on the coupling strength and the time duration $\delta t$ of the system--environment interaction because they are matrix elements of the time-evolution operator:
\begin{align}
a &= {}_e\bra{\sigma} _s\bra{P} \; U(\delta t) \; \ket{S}_s \ket{0}_e,	\\
b &= {}_e\bra{0} _s\bra{S} \; U(\delta t) \; \ket{S}_s \ket{0}_e. \label{eq:def_b}
\end{align}
This form of open system dynamics is generated by an interaction Hamiltonian of the generic form
\beq \label{eq:Hdecay}
H_\text{int} = g \Big( \ket{P}_s\bra{S}\otimes\ket{\sigma}_e\bra{0} + \ket{S}_s\bra{P}\otimes\ket{0}_e\bra{\sigma} \Big),
\eeq
where $g$ denotes the coupling strength.
In the present work, we will only focus on the generic structure of the interaction Hamiltonian. For the discussion of specific examples that are relevant in the radical pair model see~\citep{Ilichov,Ivanov,Shushin}.

As a preparation for what follows, we need an estimation of $b$ as given in~\eqref{eq:def_b}. By expanding $U(\delta t)$ to second order, and using the Hamiltonian~\eqref{eq:Hdecay}, we find that
\beq \label{eq:bApprox}
b \approx 1 - \frac{1}{2}\; {}_e\bra{0} _s\bra{S} \;H_\text{int}^2 \; \ket{S}_s \ket{0}_e \; \delta t^2
= 1-\epsilon,
\eeq
the second term of which we denote by $\epsilon$. Within this approximation $\epsilon$ is a small positive quantity, and hence $b$ is real.

According to the dynamical map~\eqref{eq:decayS} and \eqref{eq:decayT}, an arbitrary pure state that starts in the singlet/triplet subspace, i.e.\ with $\abs{c_S}^2+\abs{c_T}^2=1$, would evolve as follows:
\beq
\label{eq:decayPure}
\Big( c_S \ket{S}_s + c_T \ket{T}_s \Big)\ket{0}_e
\to
\Big( b\, c_S \ket{S}_s + c_T \ket{T}_s \Big) \ket{0}_e
+ a\, c_S \ket{P}_s \ket{\sigma}_e.
\eeq
It is the properties of the combined state of system and environment that is important for the understanding of the intrinsic dynamics of the system.
According to the axioms of quantum theory, a measurement that provides information of whether or not the reaction (with product state $\ket{P}_s$) occurred in the system, yields a positive outcome with probability $\abs{a\, c_S}^2$ and result in the measured state $\ket{P}_s \ket{\sigma}_e$ of system and environment.
A negative measurement outcome would simply return the (appropriately renormalized) state in the singlet/triplet subspace with the environment in $\ket{0}_e$.
Since here a system state in the singlet/triplet manifold only appears together with environment state $\ket{0}_e$, and the product state only with $\ket{\sigma}_e$, a measurement of the environment state instead of the system state would yield the same results.

When only the system degrees of freedom are considered, it is necessary to trace over the environment degrees of freedom.
Tracing out environmental degrees of freedom is formally equivalent to the already mentioned ``unread projective measurement'' interpretation.
That is, a projective (von Neumann type) measurement of the environment state, e.g.\ of state $\ket{0}_e$ or $\ket{\sigma}_e$, also collapses the system into its respective state with corresponding probabilities, as given according to~\eqref{eq:decayPure}, for example.
Ignoring the measurement outcomes is described by a mixed state composed of the individual states of the measurement outcomes, weighted by their respective probabilities of occurrence.
The same result is obtained by the partial trace over the environment.
A priori, this description applies to the quantum state of a \emph{single} radical pair, i.e.\ the proper state of a single radical pair (molecule) is given by a mixed density matrix.
In the case of ensembles of many radical pairs, measurement probabilities for the single radical pair translate into frequencies of occurrence when the same measurement is carried out on all members of the ensemble, and provided that there are no correlations between different radical pairs of the ensemble, and that the state of all radical pairs are identical copies of each other. Only if the latter assumptions hold, the density matrix of the single radical pair provides a complete description of the ensemble.

In the present context, it is of primary interest how the state evolves in the singlet/triplet subspace, and  thus in addition to tracing out the environment, one projects the system onto this singlet/triplet subspace.
The projection will in general yield a state $\rho$ in the singlet/triplet subspace that is not normalized.
The procedure of normalizing this state yields the (normalized) reduced state of the system that is produced by a measurement of whether or not a reaction product $\ket{P}_e$ occurred, in the case of a negative outcome.
The fraction of the state that started in an arbitrary initial pure state and has not yet undergone a recombination reaction is therefore described by the following mapping:
\beq
\rho(0) =
\begin{pmatrix} \abs{c_S}^2 & c_S\, c_T^* \\ c_S^*\, c_T & \abs{c_T}^2 \end{pmatrix}
\to
\frac{1}{1-\abs{a\, c_S}^2}
\begin{pmatrix} \abs{b}^2 \abs{c_S}^2 & b\, c_S\, c_T^* \\ b^*\, c_S^*\, c_T & \abs{c_T}^2 \end{pmatrix}
= \rho(\delta t).
\eeq
Note that, as an obvious consequence of \eqref{eq:decayPure}, a factor $b$  appears only in combination with factors $c_S$ .

An arbitrary, possibly mixed initial state of the electronic spin system will therefore obey the nonlinear, conditional evolution
\beq
\label{eq:decayMixed}
\rho(0) = \begin{pmatrix} \rho_{SS} & \rho_{ST} \\ \rho_{TS} & \rho_{TT} \end{pmatrix}
\to
\frac{1}{1-\abs{a}^2\rho_{SS}}
\begin{pmatrix} \abs{b}^2 \rho_{SS} & b\, \rho_{ST} \\ b^* \rho_{TS} & \rho_{TT} \end{pmatrix}
= \rho(\delta t),
\eeq
where the final state is normalized by the probability $(1-\abs{a}^2 \rho_{SS})$ of \emph{not} having reacted.

In case of a \emph{single} RP, this evolution would correspond to the evolution of the RP under the condition that it has not yet recombined.
In case of \emph{uncorrelated ensembles} of RPs, this translates into the subensemble of molecules that are still present (have not yet recombined) and whose absolute number scales with the normalization factor.

\subsection{Quantum measurement-induced pure dephasing}

A scenario different from spontaneous decay dynamics is given by a system--environment interaction without a recombination of the RP spins but where information regarding the singlet/triplet character of the RP spin is left in the environment, i.e.\ the spin character is effectively \emph{measured} by the environment.
In such a measurement interaction, the environment changes its state, depending on the RP spin character in a different way, which is captured by the following dynamical mapping for a time step $\delta t$:
\begin{align}
\ket{S}_s \ket{0}_e & \to a_S \ket{S}_s \ket{\sigma}_e + b_S \ket{S}_s \ket{0}_e, \label{eq:dephS} \\
\ket{T}_s \ket{0}_e & \to a_T \ket{T}_s \ket{\tau}_e + b_T \ket{T}_s \ket{0}_e. \label{eq:dephT}
\end{align}
This dynamics is generated by an interaction Hamiltonian of the form
\beq \label{eq:Hdeph}
H_\text{int}
= g_S \ket{S}_{s}\bra{S} \otimes \Big( \ket{\sigma}_{e}\bra{0} + \ket{0}_{e}\bra{\sigma} \Big)
+ g_T \ket{T}_{s}\bra{T} \otimes \Big( \ket{\tau}_{e}\bra{0} + \ket{0}_{e}\bra{\tau} \Big),
\eeq
where $g_S$ and $g_T$ are the respective interaction strengths for the environment coupling to the singlet/triplet character.

In contrast to the decay process in \eqref{eq:decayS} and \eqref{eq:decayT}, the present mapping contains the additional term $a_T \ket{T}_s \ket{\tau}_e$ in \eqref{eq:dephT} for the triplet, and there is no state $\ket{P}_s$.
Although the singlet and triplet components of the system do not seem to be affected, as the spin part of the composite final state is again a singlet or triplet with states $\ket{S}_s \big(a_S \ket{\sigma}_e + b_S \ket{0}_e \big)$ and $\ket{T}_s \big(a_T \ket{\tau}_e + b_T \ket{0}_e \big)$, respectively, the environment \emph{does} change conditioned on the spin character and thereby also affects the spin state.
With probabilities $\abs{a_S}^2$ and $\abs{a_T}^2$ the environment senses the presence of the singlet and triplet, respectively, i.e.\ in particular the triplet character, which differs from the situation of the spontaneous decay dynamics.

Thus, for the present interaction, the environment effectively performs a measurement of the spin because it evolves conditioned on the spin character and thereby effectively senses the presence of a singlet or triplet.
A measurement of the environment state would yield the respective state of the spin system, i.e.\ $\ket{S}_s$ or $\ket{T}_s$ for states $\ket{\sigma}_e$ and $\ket{\tau}_e$ of the environment, respectively, or an undetermined state for $\ket{0}_e$. 
Accordingly, the environment states $\ket{\sigma}_e$ and $\ket{\tau}_e$ constitute the measurement outcomes of this singlet/triplet measurement, and thus play the role of pointers of a measurement device that indicates the presence of a singlet or triplet.
Given a singlet or triplet state, this measurement produces a conclusive outcome with probabilities $\abs{a_S}^2$ and $\abs{a_T}^2$, respectively.
With the remaining probability $\abs{b_S}^2$ and $\abs{b_T}^2$, the singlet or triplet state, respectively, is not detected, and the environment (measurement device) remains in its initial (neutral) state~$\ket{0}_e$, which thus represents an inconclusive measurement outcome.

Since during this dynamics, the singlet fraction in the subspace of the unrecombined radical pair does not change, -- it is not taken out in the form of reaction products, and moved to the $\ket{P}_s$ subspace -- the singlet population is not affected by this dynamics.
To simplify the following discussion, let us assume equal measurement probabilities (``detector efficiencies'') for singlet and triplet.
That is, we assume $b_S=b_T \equiv b$ and thus $a_S=a_T \equiv a$, and can therefore drop the index.
When restricting our consideration to the spin system, as done for the decay scenario by means of tracing out the environment, the resulting spin state is a probabilistic mixture of the detected singlet/triplet fractions $\ket{S}_s\bra{S}$ and $\ket{T}_s\bra{T}$, each with probability $\abs{a}^2$, and the undetected (coherent superposition) spin state that occurs with probability $\abs{b}^2=1-\abs{a}^2$.
Therefore, only the singlet/triplet coherences are destroyed during the evolution:
\beq
\label{eq:dephMixed}
\rho(0) = \begin{pmatrix} \rho_{SS} & \rho_{ST} \\ \rho_{TS} & \rho_{TT} \end{pmatrix}
\to
\begin{pmatrix} \rho_{SS} & \abs{b}^2 \rho_{ST} \\ \abs{b}^2 \rho_{TS} & \rho_{TT} \end{pmatrix}
= \rho(\delta t).
\eeq
When compared to the decay scenario~\eqref{eq:decayMixed}, the state automatically remains normalized because it completely remains in the singlet/triplet manifold and, as a formal consequence, the factor $\abs{b}^2$ in the singlet population disappears because here it is accompanied by the contribution $\abs{a}^2$ of the detected singlet that is not taken to the reaction product state.

This quantum measurement dynamics results in pure dephasing, but it does not capture the transformation of the singlet into a reaction product.
As such, pure dephasing might well appear as an additional process that takes place in the radical pair model and happens in parallel, but it cannot represent the \emph{recombination} kinematics~\cite{Shushin}.

\subsection{Quantum measurement-induced recombination} \label{sec:MapMeas}

In contrast to the spontaneous decay dynamics and pure dephasing treated in the previous sections, an interaction between the RP spin system and the environment is conceivable that involves two steps.
First, the RP spin character is measured by the environment, and second, dependent on the measurement outcome, the detected singlet fraction is transformed into a reaction product.
This can be accounted for by combining the dynamical maps of pure dephasing and spontaneous decay into a \emph{quantum measurement-induced recombination} map:
\begin{align}
\ket{S}_s \ket{0}_e & \to a_S \ket{P}_s \ket{\sigma}_e + b_S \ket{S}_s \ket{0}_e, \label{eq:measS} \\
\ket{T}_s \ket{0}_e & \to a_T \ket{T}_s \ket{\tau}_e + b_T \ket{T}_s \ket{0}_e. \label{eq:measT}
\end{align}
Here, we assumed that upon detection, the singlet is transformed with unit efficiency into reaction products and hence there is no term of the form $\ket{S}_s\ket{\sigma}_e$.
We emphasize that the fact, that the singlet fraction is transported into a reaction product state $\ket{P}_s$, after the environment has sensed the singlet, is rather irrelevant to the result that information about the singlet character is left in the environment in the form of the contribution~$\ket{\sigma}_e$.
Note, that although we use the same notation for the states of the environment to indicate their  correlation to the radical pair's spin state, here the character of the environmental state differs drastically from the case in \eqref{eq:dephS} since product formation implies a change in the electronic orbital degree of freedom. Often during the product formation energy is released. One can thus generally expect that the changes in the environment due to the present type of measurement involve larger energy scales or reorganization processes than the dephasing treated in the previous section.

To keep the following discussion transparent, let us again simplify the present scenario by assuming equal measurement probabilities (efficiencies) for singlet and triplet. That is, we assume $b_S=b_T \equiv b$ and thus $a_S=a_T \equiv a$.
Then, an arbitrary pure initial spin state evolves according to
\beq
\Big( c_S \ket{S}_s + c_T \ket{T}_s \Big) \ket{0}_e
\to
a \Big( c_S \ket{P}_s\ket{\sigma}_e + c_T \ket{T}_s\ket{\tau}_e \Big)
+ b \Big( c_S \ket{S}_s + c_T \ket{T}_s \Big) \ket{0}_e.
\eeq
Thus, with probability $\abs{a}^2$, singlet and triplet are detected by the environment, which changes its state correspondingly, but in addition, the detected singlet fraction is transformed into the reaction product state $\ket{P}_s$.
Again, with remaining probability $\abs{b}^2$, initial spin state and environment remain unaffected.

Please note, that in the (generic) case that the measurement probabilities are not equal, the spin state changes even if the state is not detected by the environment (because $b_S \neq b_T$ cannot be factored out as is done with $b$ in the equation above).
Therefore, the state evolution under unsuccessful measurements by the environment will also have physically observable consequences!

By tracing out the environmental degrees of freedom and projecting to the singlet/triplet manifold, the evolution of the density matrix of the initial spin state in the singlet/triplet subspace is described as
\beq
\begin{pmatrix} \abs{c_S}^2 & c_S\, c_T^* \\ c_S^*\, c_T & \abs{c_T}^2 \end{pmatrix}
\to
\frac{1}{1-\abs{a\, c_S}^2}
\begin{pmatrix} \abs{b}^2 \abs{c_S}^2 & \abs{b}^2 c_S\, c_T^* \\ \abs{b}^2 c_S^*\, c_T & \abs{c_T}^2 \end{pmatrix},
\eeq
where the final state appears only with probability $1-\abs{a\, c_S}^2$ of the state not having decayed.
This is the normalization factor.
Note that in contrast to the spontaneous decay scenario in~\eqref{eq:decayPure}, where a factor $b$ was associated with every occurrence of $c_S$, now a factor $b$ is also associated with every occurrence of $c_T$.
In the triplet population, the term $\abs{a}^2+\abs{b}^2=1$ appears as a prefactor.
An arbitrary mixed initial state of the electronic spin system thus evolves according to
\beq
\label{eq:measMixed}
\rho(0) = \begin{pmatrix} \rho_{SS} & \rho_{ST} \\ \rho_{TS} & \rho_{TT} \end{pmatrix}
\to
\frac{1}{1-\abs{a}^2\rho_{SS}}
\begin{pmatrix} \abs{b}^2 \rho_{SS} & \abs{b}^2 \rho_{ST} \\ \abs{b}^2 \rho_{TS} & \rho_{TT} \end{pmatrix}
= \rho(\delta t),
\eeq
where, again, the final state is normalized appropriately, but here by a factor $1-\abs{a}^2 \rho_{SS}$.
This equation captures the conditional evolution of the uncorrelated subensemble of radical pairs that have \emph{not} yet recombined during the time step~$\delta t$.

A comparison between the spontaneous decay scenario~\eqref{eq:decayMixed} and the quantum measurement-induced recombination scenario~\eqref{eq:measMixed} shows that, at the level of the spin system, the only difference is that the coherences in the latter scenario carry the factor $\abs{b}^2$ instead of $b$ or its complex conjugate.
However, from the viewpoint of what happens to the environment, the underlying physics differs drastically.
The mapping for the present reaction kinetics in~\eqref{eq:measS} and \eqref{eq:measT} is generated by an interaction Hamiltonian of the form
\beq \label{eq:Hmeas}
H_\text{int}
= g_S \Big( \ket{P}_s\bra{S} \otimes \ket{\sigma}_e\bra{0} + \ket{S}_s\bra{P} \otimes \ket{0}_e\bra{\sigma} \Big)
+ g_T \ket{T}_s\bra{T} \otimes \Big( \ket{\tau}_e\bra{0} + \ket{0}_e\bra{\tau} \Big).
\eeq
Compared to the interaction Hamiltonian of the spontaneous decay, the difference in comparison to the present case is what happens to/due to the \emph{triplet} character of the state!
On one hand, the interaction of the triplet with the environment induces additional dephasing in the reduced state of the spin system, on the other hand, the environment state is changed due to the presence of the triplet.
Since here the reaction products originate only from the singlet character of the radical pair, one could think that the reduced dynamics of the spin systems is only affected by the singlet character of the state.
However, in addition, the open-system dynamics of the triplet, which does not recombine, influences the state dynamics.
From the viewpoint of the mappings and the interaction Hamiltonian, the present quantum measurement-induced recombination dynamics can thus be seen as the combination of a spontaneous decay and a pure dephasing.

\bigskip

\paragraph*{}
Every conceivable interaction between the system and an environment can be represented in terms of certain general types of evolution mappings (so-called completely positive maps, where one assumes that the initial state is given as a product state between the system and the environment)~\cite{Kraus}, out of which we have chosen three relevant examples.
In addition to the mappings described above, alternative and more complicated recombination dynamics can be constructed, which may capture more effects and would include more states of the environment.
Ultimately, it remains to the experiments to distinguish, which physical degrees of freedom are relevant and which roles they play for the recombination dynamics.

An important lesson that can be taken from the presented examples is that an alternative way of distinguishing spontaneous decay reaction kinetics from quantum measurement-induced reaction kinetics, is not only by the rate at which coherences of the RP spin state evolve, but by the traces that have been left in the environment.

\section{Master equations}

From the interaction between system and environment that is described in the previous section by a unitary mapping for a single time step $\delta t$, one can derive, using a coarse graining in time and additional assumptions, master equations of Lindblad form.\cite{GardinerZoller,BreuerPetruccione}
Such master equations can also be obtained within the original Bloch-Redfield approach to describe magnetic relaxations\cite{Bloch,Redfield}.
The general form of a Lindblad-type master equation is given by
\beq \label{eq:Lindblad}
\dot\rho = \sum_i k_i \left( L_i \rho L_i^\dag -\frac{1}{2} \left(L_i^\dag L_i \rho + \rho L_i^\dag L_i \right) \right)
\eeq
for the incoherent part of the evolution. The operators $L_i$ are called Lindblad or \emph{jump} operators, and are non-hermitian in general. The $k_i$ are rates and are positive.
(Since we focus on the system--environment interaction only, the coherent part of evolution due to the system Hamiltonian $-i[H_s,\rho]$ is omitted, which effectively amounts to using an interaction picture or simply setting $H_s=0$.)
This form of evolution equation guarantees that the density matrix remains normalized and positive. The latter means that the probabilities (eigenvalues of the density matrix) are non-negative.
The master equation is time-local, i.e.\ $\dot\rho(t)$ depends only on $\rho(t)$, and therefore cannot incorporate information of the past into the evolution.
In a rigorous derivation, this memoryless evolution results from the Markov approximation, where a time integration over environment correlation functions for earlier times is replaced by the properties of the environment at the current time.

If one wants to attribute a functional interpretation to the individual terms in the Lindblad master equation, the first term in~\eqref{eq:Lindblad} can be interpreted as the state change due to a ``successful'' interaction with the environment, whereas the second term corresponds to the state change due to an ``unsuccessful'' interaction event, i.e.\ the part of the state that remains unchanged due to the system--environment interaction but is renormalized by the corresponding probability and thereby changed.
In the present context, the first term will be associated to the recombination kinematics that transforms the radical pair to the reaction product, and thereby takes population out of the singlet/triplet manifold.
The second term leaves the populations of the radical pair spin characters invariant, but modifies the conditioned evolution, which thus exhibits a non-unitary component.

The mappings introduced in the previous section capture the action of the time evolution operator $U$ for a short time $\delta t$.
The expressions on the right hand side of these mappings thus represent the result of $U(\delta t)$ applied to the initial state, e.g.\ for a pure state:
\begin{align}
\ket{\Psi(\delta t)}_{se}
&= U(\delta t) \ket{\Psi(0)}_{se} \\
&= \exp( -i H_\text{int} \delta t) \ket{\psi_0}_s \ket{0}_e.
\end{align}
Each repeated application of the mapping transports the state forward in time by an additional time increment~$\delta t$.
If we assume that in every application of the mapping, the system interacts with a fresh initial state of the environment $\ket{0}_e$, we neglect memory effects of the environment due to previous interactions, i.e.\ we perform the Markov approximation.
This assumption is often reasonable if there are sufficiently many environmental degrees of freedom available, and/or the internal dynamics of the environment is fast enough such that correlations decay quickly and the environment effectively returns to its previous state, e.g.\ a photon or heat released into the environment quickly propagates away and does not come back.

In order to compare the dynamics on the level of master equations, rather than on the level of dynamical mappings, we will now discuss the form of the Lindblad master equations arising for the three paradigmatic cases of reaction kinematics presented in the previous section.

\subsection{Spontaneous decay}

From the dynamical map of the spontaneous decay kinematics, we can derive the differential evolution equation for the density matrix of the system.
Instead of deriving this master equation directly from the system-environment interaction Hamiltonian, we employ the dynamical maps and provide a sketch of how to relate them to the master equations of Lindblad form. For detailed derivations of Lindblad master equations from system-environment interactions see Refs.~\cite{GardinerZoller,BreuerPetruccione}, for example.

We start with the unitary map applied to a general system state as given in~\eqref{eq:decayPure} and trace over environment degrees of freedom:
\beq
\rho(0) \to \rho(\delta t)
= \abs{a}^2 \ket{P}\bra{S} \rho(0) \ket{S}\bra{P}
+ \Big( b \proj{S} + \proj{T} \Big) \rho(0) \Big( b^* \proj{S} + \proj{T} \Big).
\eeq
Since $\rho(0)$ is assumed to have only components in the singlet/triplet manifold, expressions of the form $\proj{T}\rho(0)$ can be replaced with $\rho(0)-\proj{S}\rho(0)$ and we obtain, using the notation $Q_S=\proj{S}$ and $Q_T=\proj{T}$ for the projectors:
\beq \label{eq:drhoDecay}
\rho(\delta t) =
\rho(0) + \abs{a}^2 \ket{P}\bra{S} \rho(0) \ket{S}\bra{P}
+ \abs{b-1}^2 Q_S \rho(0) Q_S + (b-1) Q_S \rho(0) + (b^*-1) \rho(0) Q_S. 
\eeq
Since, for short times $\delta t$, we can write $b=1-\epsilon$ with a small real quantity $\epsilon$ (cf.\ equation~\eqref{eq:bApprox}), it follows from $\abs{a}^2+\abs{b}^2=1$ that $\abs{a}^2 \approx 2\epsilon$ in leading order.
With these relations, and keeping only terms linear in~$\epsilon$, equation~\eqref{eq:drhoDecay} is reduced to:
\beq
\delta\rho = \rho(\delta t)-\rho(0) \approx \abs{a}^2 \ket{P}\bra{S} \rho(0) \ket{S}\bra{P}
+ (b-1) \Big( Q_{S} \rho(0) + \rho(0) Q_{S} \Big).
\eeq
Comparing this with the general expression for the Lindblad master equation~\eqref{eq:Lindblad}, we see that there is only one Lindblad operator, with the following assignments:
\begin{align}
L &= \ket{P}\bra{S}, \\
L^\dag L &= \ket{S}\bracket{P}{P}\bra{S} = \proj{S} = Q_S,	\qquad \text{and} \label{eq:LdagL}\\
\abs{a}^2 &= k \delta t
\end{align}
which is consistent with $\abs{a}^2+\abs{b}^2=1$ and hence $b\approx 1-\frac{1}{2}\abs{a}^2$ for small $\delta t$.
The resulting Lindblad operator is also suggested by the interaction Hamiltonian~\eqref{eq:Hdecay} because it induces a ``quantum jump'', i.e.\ a change of state from the singlet spin component to the reaction product state with rate $k$, and thereby yields the standard master equation for spontaneous decay dynamics~\cite{QuantumOptics}.
For example, for the case of an electromagnetic field mode in a leaky cavity, $L$ would simply be the corresponding photon annihilation operator, while for an excited atom that interacts with the electromagnetic vacuum, $L$ would be the atomic lowering operator.
In the context of the radical pair mechanism, Gauger et al.~\cite{Gauger} have recently employed this form of a spontaneous decay to model the recombination dynamics.
With the expression for $L$ at hand, the master equation~\eqref{eq:Lindblad} takes the form
\beq
\dot\rho = k\proj{P} \rho_{SS} -\frac{1}{2} k \Big( Q_{S}\rho + \rho Q_{S} \Big).
\eeq
After performing a projection to the subspace of the singlet/triplet component, the first term vanishes, and the master equation for the reaction operator reads 
\beq
[\dot\rho]_\text{rxn} = -\frac{1}{2} k \Big( Q_S \rho + \rho Q_S \Big),
\eeq
which is just the Haberkorn form of the recombination dynamics in~\eqref{eq:Haberkorn}.
Since the first term of the Lindblad equation is missing, it can no longer preserve the trace of the density matrix within the singlet/triplet manifold.
In matrix form, the equation reads
\beq \label{eq:MEdecayMatrix}
\frac{\partial}{\partial t} \begin{pmatrix}
\rho_{SS} & \rho_{ST} \\
\rho_{TS} & \rho_{TT} \end{pmatrix}
=
\begin{pmatrix}
-k \rho_{SS} & -\frac{1}{2}k\rho_{ST} \\
-\frac{1}{2}k\rho_{TS} & 0
\end{pmatrix}.
\eeq
It shows that the coherences in the singlet/triplet basis decay at half the rate of the singlet population.
A decay out of the triplet manifold can be included in an analogous way.

At this point a short remark about coherences and their decay is in order.
In the context of radical pairs, coherences are always considered in the basis of singlet and triplet, i.e.\ matrix elements of the kind $\bra{S}\rho\ket{T}$.
Note, that in other fields coherences in a quantum system are usually considered in a product basis of its constituents.
The singlet/triplet basis states are eigenstates of a collective property of the \emph{coupled} spin pair, whereas the \emph{product basis} $\{\ket{\up\up},\ket{\up\down},\ket{\down\up},\ket{\down\down}\}$ is represented by the outer products of basis states of the individual (uncoupled) spins.
The loss of coherences (dephasing) in the singlet/triplet basis, for instance due to a singlet/triplet measurement, is thus not equivalent to a dephasing in the product basis, i.e.\ the decay of a matrix element of the kind $\bra{\up\down}\rho\ket{\down\up}$.
The latter emerges due to \emph{local} measurements of the individual spins, for example.
Whereas coherence of the former type represents the presence of (electron) spin polarization in spin chemistry, i.e. a preponderance of $\ket{\up}$ or $\ket{\down}$ electron spin on the individual radical species, coherence of the latter type goes along with the concept of entanglement in quantum physics.

\subsection{Quantum measurement-induced pure dephasing}

In a similar fashion as it was done in the case of spontaneous decay dynamics, one can construct the relation between the unitary mapping and the master equation for the pure dephasing dynamics.
For dephasing in the singlet/triplet basis, a corresponding argument leads to the two Lindblad operators
\beq
L_S= \ket{S}\bra{S}=Q_S
\qquad \text{and} \qquad
L_T=\ket{T}\bra{T}=Q_T,
\eeq
i.e.\ the measurement operators of the singlet and triplet state.
With respective rates $\tilde{k}_S=\abs{a_S}^2/\delta t$ and $\tilde{k}_T=\abs{a_T}^2/\delta t$, the Lindblad equation then reads
\begin{align}
\left[\dot \rho\right]_\text{rxn}
&= \sum_{i=S,T} \tilde{k}_i \left( Q_i\rho Q_i -\frac{1}{2}\big( Q_i\rho + \rho Q_i \big) \right), \\
&= (\tilde{k}_S + \tilde{k}_T) \left( Q_S\rho Q_S -\frac{1}{2}\big( Q_S\rho + \rho Q_S \big) \right),
\end{align}
where for the second step we used that $Q_S+Q_T=\one$ in the singlet/triplet subspace.
According to this master equation, the density matrix evolves as
\beq
\frac{\partial}{\partial t} \begin{pmatrix}
\rho_{SS} & \rho_{ST} \\
\rho_{TS} & \rho_{TT} \end{pmatrix}
=
\begin{pmatrix}
0 & -\frac{1}{2}(\tilde{k}_S+\tilde{k}_T) \rho_{ST}\\
-\frac{1}{2}(\tilde{k}_S+\tilde{k}_T) \rho_{TS} & 0
\end{pmatrix}.
\eeq
Here, only the coherences decay with rate $(\tilde{k}_S+\tilde{k}_T)/2$.
When compared to the dephasing master equation~\eqref{eq:Kominis}, this rate amounts to
\beq
\tilde{k}_S+\tilde{k}_T = k_S,
\eeq
i.e.\ the effective rate appearing in~\eqref{eq:Kominis}.

Regarding the additional \emph{ad hoc} expressions for the formation of reaction products as introduced by Kominis in various ways~\cite{Kominis,Kominis2011}, it is unclear whether a microscopic physical model with a system--environment interaction exists from which these effective, phenomenological master equations can be derived.

\subsection{Quantum measurement-induced recombination dynamics}

Finally, we construct the Lindblad-type master equation for a quantum measurement-induced recombination dynamics.
From the infinitesimal change of state in the system, which we obtain from the mapping~\eqref{eq:measMixed} in the full space, we arrive at the following equation when proceeding in a similar way as explained above:
\begin{align}
\delta\rho = \rho(\delta t)-\rho(0) &\approx
\abs{a_S}^2 \ket{P}\bra{S}\rho(0)\ket{S}\bra{P}
+ (b_S-1) \Big(\proj{S}\rho(0)+\rho(0)\proj{S}\Big) \\
&+ \abs{a_T}^2 \proj{T}\rho(0)\proj{T}
+ (b_T-1) \Big(\proj{T}\rho(0)+\rho(0)\proj{T}\Big). \nonumber
\end{align}
We can now identify two Lindblad operators, which are also suggested by the interaction Hamiltonian~\eqref{eq:Hmeas}:
\begin{align}
L_S &= \ket{P}\bra{S} & \text{with}&  & \abs{a_S}^2 &= k_S \delta t & \text{and}& \\
L_T &= \ket{T}\bra{T} = Q_T & \text{with}&  & \abs{a_T}^2 &= k_T \delta t.
\end{align}
After projecting again to the singlet/triplet subspace, the master equation no longer contains the term $L_S\rho L_S^\dag$ and, by using $L_S^\dag L_S=Q_S$ as in~\eqref{eq:LdagL}, it takes the form
\begin{align}
\left[\dot\rho\right]_\text{rxn}
&= -\frac{1}{2} k_S \Big( Q_S \rho + \rho Q_S \Big)
+ k_T \left(Q_T \rho Q_T -\frac{1}{2}\Big( Q_T \rho + \rho Q_T \Big) \right) \\
&= -\frac{1}{2} k_S \Big( Q_S \rho + \rho Q_S \Big)
+ k_T Q_S \rho Q_S -\frac{1}{2} k_T \Big( Q_S \rho + \rho Q_S \Big) \\
&= -\frac{k_S+k_T}{2} \Big( Q_S \rho + \rho Q_S \Big)
+ k_T Q_S \rho Q_S .
\end{align}
For the special case when the singlet and triplet component are measured at the same rate (i.e.\ through a mechanism with the same efficiency), $k_S=k_T$, the second line reduces to the Jones-Hore master equation~\eqref{eq:JonesHore}.
The individual components of the density matrix now evolve according to
\beq
\frac{\partial}{\partial t} \begin{pmatrix}
\rho_{SS} & \rho_{ST} \\
\rho_{TS} & \rho_{TT} \end{pmatrix}
=
\begin{pmatrix}
-k_S \rho_{SS} & -\frac{1}{2}(k_S+k_T)\rho_{ST} \\
-\frac{1}{2}(k_S+k_T)\rho_{TS} & 0
\end{pmatrix},
\eeq
where in the case $k_S=k_T$ the coherences decay at the same rate as the singlet population, i.e.\ at twice the rate of the spontaneous decay in eq.~\eqref{eq:MEdecayMatrix}.

Looking at the underlying physical processes, i.e.\ at the full space, we see again that the relevant difference between the spontaneous decay scenario and the measurement-induced recombination is the dynamics that affects the triplet state.
It relates to the information about the triplet character that is obtained by the environment through an interaction or ``unread measurement''.
For a suitable choice of rates $k_S$ and $k_T$, one obtains an entire manifold of processes that may show various decay rates of the coherences.
In the limit $k_T=0$, i.e.\ vanishing interaction of the triplet with an environment and thus no measurement of the triplet character, we obtain the Haberkorn recombination dynamics of a spontaneous decay.
In the other limit of $k_S=0$, a pure dephasing dynamics emerges.
Intermediate regimes with both $k_S>0$ and $k_T>0$ yield recombination kinematics of the kind similar to the quantum measurement-induced recombination, with the Jones-Hore kinematics as a special case for $k_S=k_T$.

\section{Discussion}

Let us finally compare the three considered models for the reaction kinetics.
It is obvious that, on the level of the system in the RP singlet/triplet subspace, three entirely different system--environment interactions create very similar time evolutions that only differ in the rate at which the coherences decay.
Observing only the RP spin system in the singlet/triplet subspace may therefore not be sufficient to decide on the generating process and the underlying physics of the recombination kinematics.
For example, by combining several independent processes affecting the radical pair at the same time, their Lindblad master equations are simply added.
Therefore, in a process where, on one hand, a spontaneous decay from the singlet at rate $k_S$ and, on the other hand, a pure dephasing (``measurement'') process with rates $\tilde{k}_S = k_S/2$ and $\tilde{k}_T = k_S/2$ affect the spin system simultaneously, the evolution can be interpreted as a quantum measurement-induced recombination with rates $k_S=k_T$ as proposed by Jones and Hore. Indeed, the master equations for the radical pair would be identical.
It is therefore not surprising that a similar phenomenology on the level of the RP spin systems can be obtained by even more complicated models as recently considered by Jones et al.~\cite{Jones2011}, which include a larger number of reaction stages than those considered here.

In order to distinguish between dynamics due to a quantum measurement-induced recombination, and a suitable combination of spontaneous decay and dephasing, it is necessary to also consider the involved degrees of freedom of the environment. Including the environment degrees of freedom into the description allows us to identify the origin and influence of ``measurements'' during the recombination dynamics. The dynamics caused by the interaction of system and environment represents a situation in which the environment effectively measures the (spin) state of the system. Since these ``measurement results'' are never read out, the single radical pair system per se is correctly described by a mixed state that evolves according to the master equation.
We thereby complement previous treatments~\cite{Ilichov,Ivanov,Shushin} focusing on the problem of identifying possible relevant interaction Hamiltonians, by embedding them into a framework that allows one to identify quantum measurements of the radical pair spin system.

By including the interaction with a (quantum) environment from the beginning, the origin of the terms in the master equation becomes transparent, and avenues emerge that allow for distinguishing the various proposed reaction mechanisms by the traces that the state of the system leaves in the environment.
Once the relevant environmental degrees of freedom are identified, their experimental examination would provide insight as to what information about the RP spin character has been left in the environment. The absence of an interaction, however, can only be stated with some uncertainty determined by the experimental precision.
Furthermore, an experimental measurement of the environment would disclose information on the system, and thereby project it into the respective system state.
It would thus become possible to \emph{unravel} the decoherence dynamics of the RP spin state, and observe the system dynamics as it stochastically evolves under the system--environment interaction.
Such an observation realizes the dynamics in the form of individual stochastic (almost) pure state trajectories instead of the averaged mixed state description provided by the master equation.
Similar processes are well-known in the field of quantum optics, where suitable experiments have been performed with systems of trapped ions~\cite{jumpsIons}, micro-masers and cavity QED-setups~\cite{jumpsCQED,micromaser}.
A translation of these or similar experiments to single molecule (i.e.\ single radical pair) systems of spin chemistry (or quantum dot systems that might mimic radical pairs) represents an experimental challenge, but would yield an unprecedented level of access to spin chemical reactions on the quantum level.
Possible experimental handles to environment degrees of freedom may include single photon detection for fluorescent radical pair recombination, the detection of dissipated heat by spectroscopy or thermal lens/thermal grid techniques, and the resolution of the intermolecular Brownian molecule motion.

In the present treatment, we have restricted ourselves to a simple ``two-level system'' for the RP spin manifold consisting of the $\ket{S}$ and $\ket{T_0}$ spin states.
This scenario is approximately realized for singlet-born radical pairs in the high-field limit, where the spin states $\ket{T_\pm}$ are energetically separated due to a large Zeemann interaction.
However, possible open system dynamics of a two-level system are limited to essentially $T_1$- and $T_2$-type processes.
For the interesting low-field regime, however, the fully accessible four-dimensional spin space offers a richer dynamical phenomenology.

\section{Conclusion}

The presented framework gives a common basis, on which the various reaction operators proposed in the literature can be equally constructed and compared.
By adopting a view of system and environment, it becomes natural to identify quantum measurements in the reaction processes, thereby demystifying the quantum measurement approach to spin chemistry.
Within the presented framework, we exemplarily derived the reaction operators proposed in the literature, while providing a systematic way of constructing new ones.
Although both the traditional as well as the ``quantum measurement''-inspired reaction operators emerge as formally correct proposals, the ultimate decision about which of the reaction operators are the appropriate model in a given scenario will certainly depend on the precise physical situation.

The open quantum system approach to the spin-dependent recombination kinetics of radical pairs presented here provides a general framework for justifying and establishing appropriate reaction operators for the stochastic Liouville equations customary in spin chemistry.
Proceeding from general evolution mappings, that specify the particular interactions between spin system and environment, through pertinent Lindblad equations in the complete system/environment space to the final stochastic Liouville equation of the reduced spin system, defines a transparent approach to the theoretical treatment of spin-dependent radical pair kinetics.
An important lesson that we can take from this discussion is the following:
We see that the rate at which the coherences between singlet and triplet decay is no longer the primary criterion to decide whether a quantum measurement of the spin character occurred.
The discussion of the correct master equation should therefore not focus on formal expressions of rates but rather include the traces of the spin character that have been left in the environment.
It is the involvement of the environment that justifies the description and treatment of radical pair kinetics in terms of quantum measurement theory.
The interpretation of the recombination kinetics in terms of quantum measurements can thus be embedded in the general context of open quantum systems.

\section{Acknowledgments}

M.T. and H.J.B. acknowledge support in part by the Austrian Science Fund (FWF): F04011, F04012.



\end{document}